\documentclass[twocolumn]{aa} 

\usepackage{natbib}
\usepackage{amsmath}
\usepackage{graphicx}
\usepackage{txfonts}
\usepackage{textcomp}

\makeatletter
\def\@biblabel#1{}
\makeatother

\newcommand{\mjup}{M$_{\rm J}$\,}
\newcommand{\msun}{M$_\odot$\,}
\newcommand{\mstar}{M$_\star$}
\newcommand{\ms}{m\,s$^{-1}$\,}
\newcommand{\teff}{T$_{{\rm eff}}$\,}
\newcommand{\mplan}{m$_b$\,sin$i$}
\renewcommand{\cite}{\citealp}

\begin{document}


\title{Giant planets around two intermediate-mass evolved stars and confirmation of the planetary nature of HIP\,67851{\it c}
\thanks{Based on observations collected at La Silla - Paranal Observatory under
programs ID's 085.C-0557, 087.C.0476, 089.C-0524, 090.C-0345 and through the Chilean Telescope Time under programs
ID's CN 12A-073, CN 12B-047, CN 13A-111, CN-14A-128 and CN-15A-48.}}

  \titlerunning{Planets around the giant stars HIP\,65891 and HIP\,107773.}
   \author{M. I. Jones \inst{1}
           \and J. S. Jenkins \inst{2}
           \and P. Rojo \inst{2}
           \and C. H. F. Melo \inst{3}}
         \institute{Department of Electrical Engineering and Center of Astro-Engineering UC, Pontificia Universidad
         Cat\'olica de Chile, Av. Vicuña Mackenna 4860, 782-0436 Macul, Santiago, Chile \\\email{mjones@aiuc.puc.cl}
         \and Departamento de Astronom\'ia, Universidad de Chile, Camino El Observatorio 1515, Las Condes, Santiago, Chile 
         \and European Southern Observatory, Casilla 19001, Santiago, Chile}

   \date{}

 
  \abstract
{Precision radial velocities are required to discover and characterize planets orbiting nearby stars.
Optical and near infrared spectra that exhibit many hundreds of absorption lines can allow the \ms  
precision levels required for such work. However, this means that studies have generally focused on solar-type dwarf stars.
After the main-sequence, intermediate-mass stars (former A-F stars) expand and rotate slower than their 
progenitors, thus thousands of narrow absorption lines appear in the optical region, permitting the 
search for planetary Doppler signals in the data for these types of stars.} 
{In 2009, we began the EXPRESS program, aimed at detecting substellar objects around evolved stars, and to study the effects of the mass and evolution 
of the host star on their orbital and physical properties.}
{We have obtained precision radial velocity measurements for the giant stars HIP\,65891 and HIP\,107773, from CHIRON and FEROS spectra. 
Also, we obtained new radial velocity epochs for the star HIP\,67851, which is known to host a planetary system. }
{We present the discovery of two giant planets around the intermediate-mass evolved star HIP\,65891 and HIP\,107773. 
The best Keplerian fit to the HIP\,65891 and HIP\,107773 radial velocities leads to the following orbital parameters: 
P=1084.5 d; \mplan = 6.0 \mjup; $e$=0.13 and P=144.3 d; \mplan = 2.0 \mjup; $e$=0.09, respectively. 
In addition, we confirm the planetary nature of the outer object 
orbiting the giant star HIP\,67851. The orbital parameters of HIP\,67851\,c are: P=2131.8\,d, m$_c$\,sin$i$ = 6.0 \mjup and $e$=0.17.}
{With masses of 2.5 \msun and 2.4 \msun, HIP\,65891 and HIP\,107773 are two of the most massive stars known to host planets. 
Additionally, HIP\,67851 is one of five giant stars that are known to host a planetary system having a close-in planet ($a <$ 0.7 AU). 
Based on the evolutionary states of those five stars, we conclude that close-in planets do exist in multiple systems around subgiants and slightly 
evolved giants stars, but probably they are subsequently destroyed by the stellar envelope during the ascent of the red giant branch phase. 
As a consequence, planetary systems with close-in objects are not found around horizontal branch stars. } 
   {}


   \maketitle
%

\section{Introduction}

Since the discovery of the first planetary system around the pulsar PSR 1257+12 (Wolszczan \& Frail \cite{WOL92}) and 
the Jupiter-mass companion to the solar-type star 51 Pegasi (Mayor \& Queloz \cite{MAY95}), 
the exoplanet field has experienced an exponential growth, leading to the discovery of $\sim$\,1200 systems\footnote{http://exoplanet.eu} 
and more than 3000 unconfirmed candidates from the {\it Kepler} mission (Borucki et al. \cite{BOR10}).  
These planetary systems have been found in very different environments and configurations, showing us that 
planetary formation is a common phenomenon in our galaxy. \newline  \indent
Although only a small fraction of exoplanets have been found around intermediate-mass stars (IMS; \mstar $\gtrsim$ 1.5 \msun), they are of great 
importance, since they allow us to understand the role of stellar mass on the orbital properties and formation efficiency,
and to test the validity of planet formation models. \newline  \indent
Bowler et al. (\cite{BOW10}) investigated the period-mass distribution of planets orbiting IMS with 
M$_\star$ $\sim$ 1.5-2.0 \msun, from a uniform sample of 31 subgiants observed by the Lick program (Johnson et al. \cite{JOH06}). 
They found that their properties are different compared  to solar-type host stars at the 4 $\sigma$ level. 
Moreover, they found that the fraction of planets orbiting those stars is $\sim$ 26\%, compared to only $\sim$ 10 \% for solar-type hosts. 
In addition, based on a much larger sample of 1266 stars observed by the California Planet Survey (Howard et al. \cite{HOW10}), 
Johnson et al. (\cite{JOHN10}) showed that there is a linear increase in the fraction of planets, from $f=0.03$ to $f=0.14$, in the mass 
range between $\sim$ 0.5 - 2.0 \msun. 
These observational results tells us that planet formation efficiency is strongly dependent on the stellar mass.
However, the reliability of the derived masses of subgiant host stars has been recently called into question.
Lloyd (\cite{LLO11,LLO13}) showed that the mass distribution of the planet-hosting subgiants is incompatible with the distribution 
derived from integrating isochrones, concluding that these stars have masses of $\sim$ 1.0-1.2 \msun. Similarly, based on Galactic 
kinematics, Schlaufman \& Winn (\cite{SCH13}) concluded that the subgiant host stars are similar in mass to solar-type host stars. 
However, if this is the case, then the planetary systems around them should exhibit the same orbital properties and detection fraction as for planets
around less massive stars. \newline \indent 
Finally, based on a sample of 373 giant stars targeted  by the Lick radial velocity (RV) survey (Frink et al. \cite{FRI02}), 
Reffert et al. (\cite{REF15}) studied the occurrence rate of planets around stars with masses between $\sim$ 1.0 - 3.0 \msun. 
They showed that the fraction of exoplanets increases with increasing 
stellar mass, with a peak at $\sim$ 1.9 \msun, and that there is a rapid drop in the occurrence rate for stars more massive than $\sim$ 
2.5 \msun. \newline \indent 
In this paper we present the discovery of two planets orbiting the intermediate-mass giant stars HIP\,65891 and HIP\,107773. These are two of the most
massive stars that are known to host substellar companions. HIP\,65891\,{\it b} and HIP\,107773\,{\it b} are the sixth and seventh substellar 
objects discovered by the EXPRESS RV survey (Jones et al. \cite{JON11,JON15}).
Additionally, we present new RV epochs of the giant star HIP\,67851. These velocities allowed us to confirm the planetary nature of the outer object
in the system, as suggested by Jones et al.  (\cite{JON15}).
The paper is organized as follows. In Sect. 2, we briefly present the observations, data reduction and the calculation methods 
to obtain the radial velocities.  In Sect. 3, we summarize the properties of the host stars. In Sect. 4, we present the orbital parameters of 
HIP\,65891\,{\it b}, HIP\,107773\,{\it b} and HIP\,67851\,{\it c}, as well as improved orbital parameters for HIP\,67851\,{\it b}. 
In Sect. 5 the stellar activity analysis is presented. Finally, in Sect. 6, we present the summary and discussion.

\section{Observations and data reduction}

We have collected a total of 26 spectra of HIP\,65891 and 36 spectra of HIP\,107773 using CHIRON (Tokovinin et al. \cite{TOK13}), a 
high-resolution stable spectrograph installed in the 1.5m telescope at Cerro Tololo Inter-American Observatory. 
Using the image slicer mode (R $\sim$ 80'000), we typically obtained a signal-to-noise-ratio (S/N) of $\sim$\,100 with 400 s of integration for 
HIP\,65891 and $\sim$\,150 s for HIP\,107773.
The data reduction was performed using the CHIRON data reduction system. The pipeline does 
a standard echelle reduction, i.e., bias subtraction, flat field correction, order tracing, extraction and wavelength calibration.
Additionally, we use an iodine cell in the `IN' position, meaning that it is placed in the light path, at the fiber exit. The cell contains
molecular iodine (I$_2$), which superimpose a forest of absorption lines in the region between $\sim$ 5000-6000 \AA\,. These lines are used as 
precise wavelength markers against which the doppler shift of the stellar spectrum is measured. The radial velocity variations were calculated 
according to the method described in Butler et al. (\cite{BUT96}) and Jones et al. (\cite{JON14,JON15}). 
We achieve a mean RV precision of $\sim$ 5 \ms from CHIRON spectra using this method. \newline \indent  
In addition, we took 24 spectra of HIP\,65891 and 28 spectra of HIP\,107773 using the Fiber-fed Extended Range Optical Spectrograph 
(FEROS; Kaufer et al. \cite{KAU99}) mounted at the 2.2m telescope at La Silla Observatory. We discarded one FEROS spectrum of HIP\,65891 since there was
a problem with a folding mirror in the calibration unit.
The typical observing time was $\sim$ 60 s and $\sim$ 180 s (for HIP\,65891 and HIP\,107773, respectively), leading to a S/N $\sim$\,100 
per pixel. The data reduction of the spectra was performed with the FEROS pipeline. 
The radial velocities were computed using the simultaneous calibration method (Baranne 
et al. \cite{BAR96}), according to the method described in Jones et al. (\cite{JON13}) and Jones \& Jenkins (\cite{JONJEN14}). 

\begin{table}
\centering
\caption{Stellar properties \label{stellar_par}}
\begin{tabular}{lrrr}
\hline\hline
\vspace{-0.3cm} \\ 
 & HIP\,65891 & HIP\,107773 & HIP\,67851 \\
\hline \vspace{-0.3cm} \\
Spectral Type         &  K0III             &   K1III             & K0III             \\
B-V (mag)             &  1.00              &   1.02              & 1.01              \\
V (mag)               &  6.75              &   5.62              & 6.17              \\
Parallax (mas)        &  7.35 $\pm$ 0.60   &   9.65  $\pm$ 0.40  & 15.16 $\pm$ 0.39  \\
\teff (K)             &  5000 $\pm$ 100    &   4945  $\pm$ 100   & 4890  $\pm$ 100   \\
L (L$_\odot$)         &  44.8 $\pm$  9.6   &   74.0  $\pm$ 9.2   & 17.55 $\pm$ 2.64  \\
log\,g (cm\,s$^{-2}$) &  2.9  $\pm$ 0.2    &   2.6   $\pm$ 0.2   & 3.15  $\pm$ 0.20  \\
{\rm [Fe/H]} (dex)    &  0.16 $\pm$ 0.10   &   0.03  $\pm$ 0.10  & 0.00  $\pm$ 0.10  \\
$v$\,sin$i$ (k\ms)    &  2.5  $\pm$ 0.9    &   2.0   $\pm$ 0.9   & 1.8   $\pm$ 0.9   \\
M$_\star$ (\msun)     &  2.50 $\pm$ 0.21   &   2.42  $\pm$ 0.27  & 1.63  $\pm$ 0.22  \\
R$_\star$ (R$_\odot$) &  8.93 $\pm$ 1.02   &   11.6  $\pm$ 1.4   & 5.92  $\pm$ 0.44  \\
\vspace{-0.3cm} \\\hline\hline
\end{tabular}
\end{table}

\section{Stellar properties}

\begin{figure}[]
\centering
\includegraphics[width=8cm,height=9cm,angle=270]{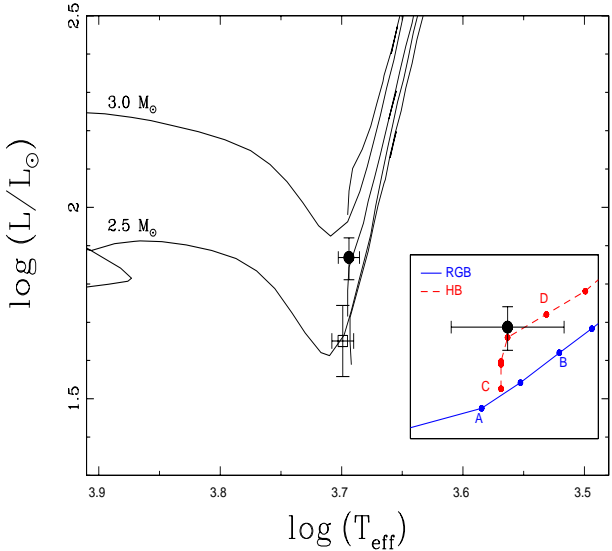}
\caption{Position of HIP\,65891 (open square) and HIP\,107773 (filled circle) in the HR diagram. Different S00 evolutionary tracks with solar 
metallicity are overplotted. The small panel shows a zoomed region of the HIP\,107773 position and its closest isomass track in the grid. 
\label{stars_position}}
\end{figure}

The stellar parameters of HIP\,65891 and HIP\,107773 are summarized in Table \ref{stellar_par}. 
The spectral types, $V$ magnitudes, $B-V$ colors and parallaxes were taken from the {\it Hipparcos} catalog (Van Leeuwen \cite{VAN07}).  
The atmospheric parameters were retrieved from Jones et al. (\cite{JON11}). 
For each star we created 100 synthetic datasets for Teff, logL and [Fe/H], assuming Gaussian distributed errors.
Then we compared these synthetic datasets with Salasnich et al. (\cite{SAL00}; S00 hereafter) models, following the method presented in 
Jones et al. (\cite{JON11}). The resulting values for M$_\star$ and R$_\star$ correspond to the mean and the RMS of the two resulting distributions. 
\newline \indent
Figure \ref{stars_position} shows a HR diagram with the positions of HIP\,65891 (open square) and HIP\,107773 (filled circle).  
For comparison, two S00 models with solar metallicity are overplotted. 
As can be seen, HIP\,65891 is most likely at the base of the Red Giant Branch (RGB) phase, since no Horizontal Branch (HB) model intersects its position. 
The small panel shows a zoomed region of the HIP\,107773 position and its closest evolutionary track in the grid (M$_\star$ = 2.5 \msun and [Fe/H] = 0.0).
The blue solid and red dashed lines correspond to the RGB and HB phase, respectively. The dots are the points in the grid.
As can be seen, it is not clear whether the star is ascending the RGB or has already reached the He-core burning phase.
However, according to these evolutionary models, the timescales between point A and B is $\sim$ 200 times shorter than between C and D.
Therefore, based on the ratio of these timescales, we conclude that HIP\,107773 is most likely a HB star. \newline \indent
The stellar properties of HIP\,67851 (retrieved from Jones et al. \cite{JON15}) are also summarized in Table \ref{stellar_par}. 

\section{Orbital parameters}

\subsection{HIP\,65891\,{\it b}}

\begin{figure}[t!]
\centering
\includegraphics[width=8cm,height=10cm,angle=270]{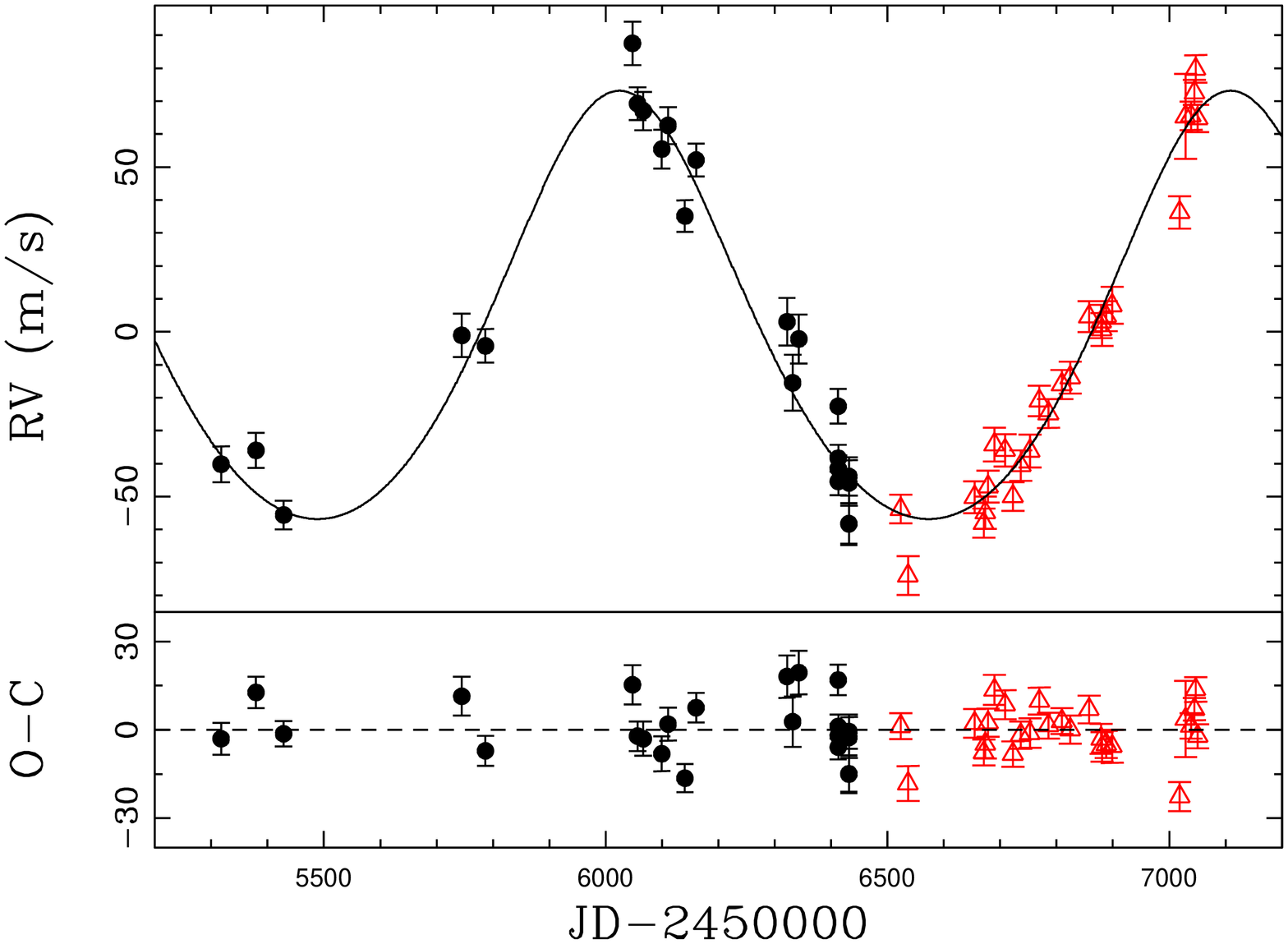}
\caption{Upper panel: Radial velocity variations of HIP\,65891. The red triangles and black circles correspond to CHIRON and FEROS velocities, respectively. 
The best Keplerian solution is overplotted (black solid line).
Lower panel: Residuals from the Keplerian fit. The RMS of the fit is 9.3 \ms.
\label{HIP65891_vels}}
\end{figure}

The RVs of HIP\,65891 are listed in Tables \ref{chiron_HIP65891_vels} and \ref{feros_HIP65891_vels}. 
A Lomb-Scargle (LS) periodogram (Scargle \cite{SCA82}) of the data revealed a strong peak around $\sim$\,1019 days with a false alarm probability (FAP) 
of $\sim$\,10$^{-7}$.  Starting from this period we use the Systemic Console 2.17 (Meschiari et al. \cite{MES09}) and we obtained a 
single-planet solution with the 
following orbital parameters: P = 1084.5 $\pm$ 21.5 d, K = 64.9 $\pm$ 2.4 \ms (corresponding to \mplan = 6.0 \mjup) and $e$ = 0.13 $\pm$ 0.05. The 
uncertainties were obtained using the bootstrap tool provided by the Systemic Console. The uncertainties in the semimajor axis and planet mass 
were computed by error propagation of these values, also including the uncertainty in the stellar mass\footnote{The Systemic Console does not include 
the contribution from the uncertainty in the stellar mass}.
The full set of orbital parameters are listed in Table \ref{HIP107773_orb_par}. Figure \ref{HIP65891_vels} shows the radial velocity curve.
The red triangles and black circles correspond to CHIRON and FEROS data, respectively. The Keplerian fit is overplotted (solid line), leading to 
a RMS of 9.3 \ms.

\subsection{HIP\,107773\,{\it b}}

\begin{figure}[t!]
\centering
\includegraphics[width=8cm,height=10cm,angle=270]{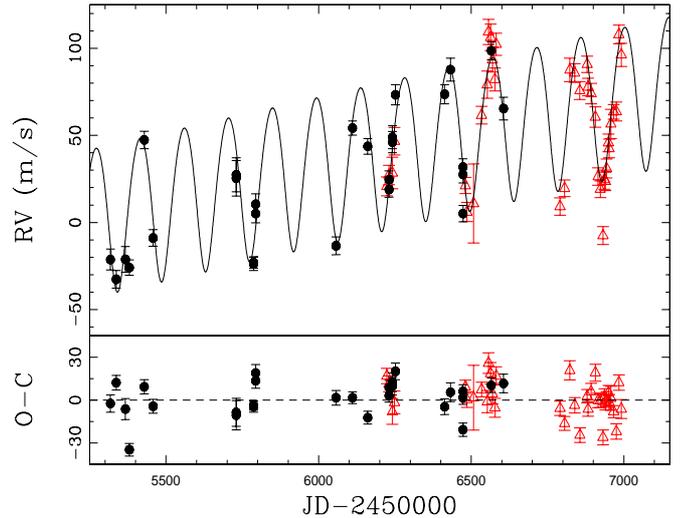}
\caption{Upper panel: Radial velocities measurements of HIP\,107773. The red triangles and black circles correspond to CHIRON and FEROS velocities, 
respectively. The best Keplerian solution, including a linear trend is overplotted (black solid line).
Lower panel: Residuals from the Keplerian fit. The RMS around the fit is 11.9 \ms.
\label{HIP107773_vels}}
\end{figure}

The radial velocity measurements of HIP\,107773 computed from CHIRON and FEROS spectra are listed in Tables \ref{chiron_HIP107773_vels} and 
\ref{feros_HIP107773_vels}, respectively. A periodogram analysis of the data revealed a 144.3-days signal with a FAP  
of $\sim$ 10$^{-7}$. The best Keplerian fit of the data leads
to a one-planet system plus a linear trend with the following parameters: P = 144.3 $\pm$ 0.4 d,\, K=42.7 $\pm$ 2.2 \ms (corresponding to 
\mplan = 2.0 \mjup), e=0.09 $\pm$ 0.08 and $\dot{\gamma}$=14.6 $\pm$ 1.8 m\,s$^{-1}$yr$^{-1}$. The solution with a linear trend significantly improves
the RMS of the fit (from 17.8 \ms to 12.0 \ms). The $F$ value computed from the ratio of the $\chi^2_{red}$ with 7 and 8 parameters is 2.1. The probability 
of exceeding such value (assuming  an $F$-distribution; see Bevington \& Robinson \cite{BEV03}) is 0.16.
From $\dot{\gamma}$ and using the relation
given in Winn et al. (\cite{WINN09}), we derived the minimum mass and orbital distance of the second planet in the system of m$_c$ $\gtrsim$ 
2.8 \mjup\, and $a_c$ $\gtrsim$ 5.9 AU. The orbital parameters are also listed in Table \ref{HIP107773_orb_par}.
Figure \ref{HIP107773_vels} shows the RV measurements from CHIRON and FEROS spectra (red triangles and black circles, respectively) and the 
Keplerian fit (solid black line). The RMS of the post-fit residuals is 11.9 \ms.   

\begin{table*}
\centering
\caption{Orbital parameters \label{HIP107773_orb_par}}
\begin{tabular}{lrrrr}
\hline\hline 
\vspace{-0.3cm} \\
                              &  HIP\,65891\,{\it b} &  HIP\,107773\,{\it b}   &  HIP\,67851\,{\it b}     & HIP\,67851\,{\it c}      \\
\hline \vspace{-0.3cm}                                                                                                       \\     
P (days)                      &  1084.5 $\pm$ 23.2   &  144.3   $\pm$  0.5     &   88.9  $\pm$ 0.1    & 2131.8  $\pm$ 88.3   \\ 
K (\ms)                       &    64.9 $\pm$ 2.4    &   42.7   $\pm$  2.7     &   45.5  $\pm$ 1.6    & 69.0    $\pm$ 3.3    \\ 
$a$ (AU)                      &    2.81 $\pm$ 0.09   &   0.72   $\pm$  0.03    &   0.46  $\pm$ 0.02   & 3.82    $\pm$ 0.23   \\
$e$                           &    0.13 $\pm$ 0.05   &   0.09   $\pm$  0.06    &   0.05  $\pm$ 0.04   & 0.17    $\pm$ 0.06   \\
m\,sin$i$ (\mjup)             &    6.00 $\pm$ 0.49   &   1.98   $\pm$  0.21    &   1.38  $\pm$ 0.15   & 5.98    $\pm$ 0.76   \\
$\omega$ (deg)                &   355.5 $\pm$ 15.5   &   166.0  $\pm$  32.6    &  138.1  $\pm$ 60.0   & 166.5   $\pm$ 20.5   \\
T$_{\rm P}$-2450000           &  6014.8 $\pm$ 49.3   &  6202.3  $\pm$  12.8    & 2997.8  $\pm$ 16.7   & 2684.1  $\pm$ 235.7  \\
$\gamma_1$ (\ms) (CHIRON)     &    7.9  $\pm$ 4.7    &   -18.4  $\pm$  4.6     &   29.0  $\pm$ 3.8    &  29.0   $\pm$ 3.8    \\
$\gamma_2$ (\ms) (FEROS)      &    2.5  $\pm$ 3.2    &    14.1  $\pm$  2.5     &  -31.5  $\pm$ 4.3    &  -31.5  $\pm$ 4.3    \\
$\dot{\gamma}$ (\ms\,yr$^-1$) &    -\,-\,-           &   14.6   $\pm$  2.5     &  -\,-\,-             &  -\,-\,-             \\
RMS (\ms)                     &    9.3               &   12.0                  &  8.9                 &  8.9                 \\
$\chi$$^2_{\rm red}$          &    3.5               &   5.7                   &  3.1                 &  3.1                 \\
\vspace{-0.3cm} \\\hline\hline
\end{tabular}
\end{table*}

\subsection{HIP\,67851\,{\it c}}

\begin{figure*}[t!]
\centering
\includegraphics[width=12cm,height=15cm,angle=270]{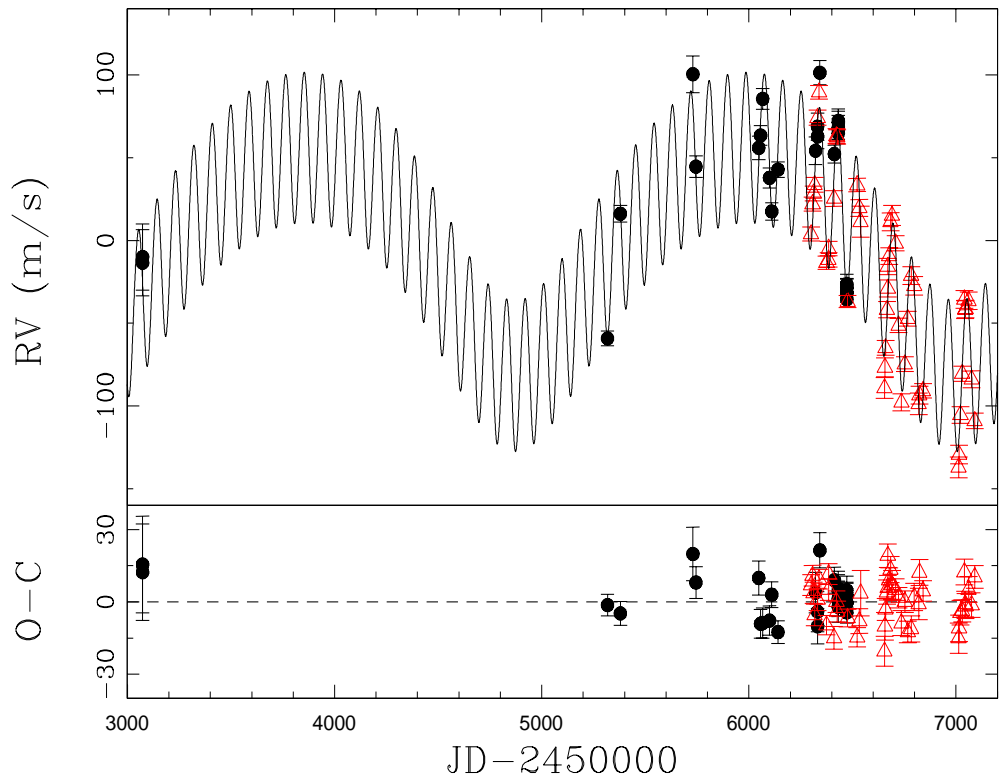}
\caption{Upper panel: Radial velocity variations of HIP\,67851. The red triangles and black circles correspond to CHIRON and FEROS velocities,
respectively. The best two-planet solution is overplotted (black solid line).
Lower panel: Residuals from the Keplerian fit. The RMS around the fit is 8.9 \ms.
\label{HIP67851_vels}}
\end{figure*}

We have obtained new CHIRON RV epochs which allowed us to fully cover the orbital period of HIP\,67851\,{\it c}, and thus to confirm its planetary
nature, as proposed by Jones et al. (\cite{JON15}). In addition, we found two FEROS spectra of HIP\,67851 in the ESO archive, that were taken in 
2004.  Figure \ref{HIP67851_vels} shows the RV curve of HIP\,67851. The red triangles and black circles, correspond to CHIRON and FEROS data, 
respectively. The best two-planets solution is overplotted (black solid line). 
The orbital parameters of HIP\,67851\,{\it c} are: P = 2131.8 $\pm$ 43.5 d, K = 69.0 $\pm$ 1.9 \ms (corresponding to
m$_c$\,sin$i$ = 6.0 \mjup) and e = 0.17 $\pm$ 0.04. These values as well as the refined orbital parameters of HIP\,67851\,{\it b} are listed in 
Table \ref{HIP107773_orb_par}. 
We note that Wittenmyer et al. (\cite{WIT15}) recently presented RV measurements of HIP\,67851 from the Pan-Pacific Planet Search (PPPS; Wittenmyer et al. 
\cite{WIT11}). They recovered the signal of HIP\,67851\,{\it b} and confirmed the presence of an outer planet in the system. However, they obtained an orbital
period of 1626 $\pm$ 26 $d$ and minimum mass of the planet of 3.6 $\pm$ 0.6 \mjup (assuming a stellar mass of 1.3 \msun), which is 
incompatible with our solution. 
The reason for this discrepancy is because their orbital solution of HIP\,67851\,{\it c} relies on one RV data-point, which is most likely an outlier.
In fact, a new reduction of the PPPS dataset, including new RV epochs, is soon to be published, and the new solution is in good agreement with the
solution presented here (Wittenmyer; private communication).

\section{Intrinsic stellar phenomena}

\begin{figure}[h!]
\centering
\includegraphics[width=7.0cm,height=8cm,angle=270]{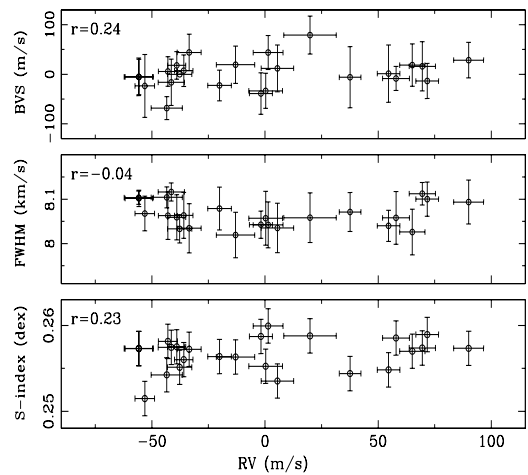}
\caption{BVS, FWHM and S-index variations versus FEROS radial velocities of HIP\,65891 (upper, middle and lower panel, respectively).\label{HIP65891_BVS}}
\end{figure}

\begin{figure}[h!]
\centering
\includegraphics[width=7.0cm,height=8cm,angle=270]{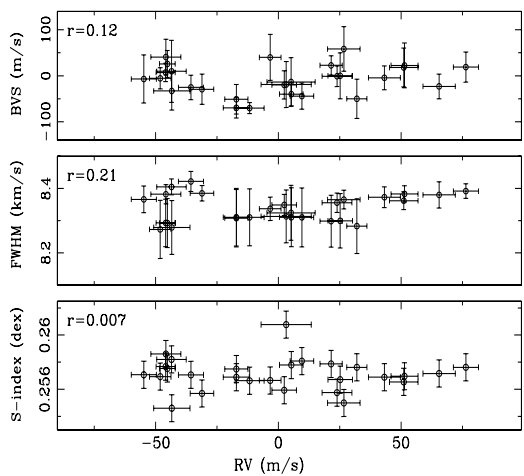}
\caption{BVS, FWHM and S-index variations versus FEROS radial velocities of HIP\,107773 (upper, middle and lower panel, 
respectively).\label{HIP107773_BVS}}
\end{figure}

To test the nature of the RV signals detected in HIP\,65891 and HIP\,107773, we analyzed the {\it Hipparcos} photometric data of these two stars.
The HIP\,65891 dataset is comprised of 123 high quality epochs, spanning 3.2 years. The photometric variability is less than 0.01 mag and a 
periodogram analysis revealed no significant periodicity in the data. Similarly, the photometric data of HIP\,107773  is comprised of 221 high 
quality measurements, with a  baseline of 3.2 years. This dataset revealed a photometric variability of $\sigma$ = 0.007 mag and no periodicity is 
observed in a LS periodogram. Also, from the projected rotational velocity and the stellar radius, we can put an upper limit on the stellar rotational 
period. In the case of HIP\,65891, we obtained a value of 179\,d, meaning that we can discard the hypothesis that the observed RV variation is related 
to the rotation of star. In the case of HIP\,107773, we computed a maximum rotational period of 293\,d, which is more than 2 times larger than RV period. 
We note that an inclination angle of $i \sim$ 29 degrees would be required so that the rotational period would match the RV period. \newline \indent
Additionally, we analyzed the line profile variations, by computing the bisector velocity span (BVS) and full width at half 
maximum (FWHM) variations of the cross-correlation function (CCF), in a similar way as presented in Jones et al. (\cite{JON14}). 
Also, we computed the chromospheric activity indexes from integrating the flux in the core of the Ca {\sc ii} HK lines,
in the same manner as described in Jenkins et al. (\cite{JEN08,JEN11}). Only FEROS spectra were used since CHIRON does not cover the spectral
region where these lines are located. These results are shown in Figures \ref{HIP65891_BVS} and \ref{HIP107773_BVS}. The Pearson correlation 
coefficients are also labelled. Clearly there is no significant correlation between these quantities and the radial velocities of any of the 
two stars. Also, a LS periodogram analysis revelead no significant peak in these quantities, with the exception of a peak (FAP $\sim$ 0.009) in the FWHM
variations of HIP\,107773, around $\sim$ 183 d. This value is significantly longer than the 144-days period observed in the RV data.\newline \indent
Lastly, most giant stars bluer than $B\,-\,V$ $<$ 1.2 exhibit pulsation-induced RV variability at the 10-20 \ms (Sato et al. \cite{SAT05}; 
Hekker et al. \cite{HEK06}), which is well below the amplitudes observed in these stars. In fact, the Kjeldsen \& Bedding  (\cite{KJE95}) scaling 
relations predict RV amplitudes of $\sim$ 4 and 7 \ms for HIP\,65891 and HIP\,107773, respectively. The corresponding lifetimes of the maximum power 
oscillations ($1/\nu_{max}$) are 2.7 and 4.7 hrs, respectively.

\section{Summary and discussion}

In this work we present precision radial velocities for the giant stars HIP\,65891 and HIP\,107773. 
The two datasets revealed periodic signals, which are most likely explained by the presence of giant planets orbiting these two stars. 
The best Keplerian fit to the HIP\,65891 data leads to the following orbital parameters: P = 1084.5 $\pm$ 21.5 d, \mplan = 6.0 $\pm$ 0.5 \ms 
and $e$ = 0.13 $\pm$ 0.05. Similarly, the orbital solution for HIP\,107773\,{\it b} is: P = 144.3 d $\pm$ 0.4, \mplan = 2.0 $\pm$ 0.2 \mjup
and $e$ = 0.09 $\pm$ 0.08 plus a linear trend of $\dot{\gamma}$=14.6 $\pm$ 1.8 m\,s$^{-1}$yr$^{-1}$.  
We derived a mass of 2.5 \msun and 2.4 \msun for HIP\,65891 and HIP\,107773, 
respectively, meaning that they are amongst the most massive stars that are known to host planets. 
Although there relatively few known planets around stars more massive than $\sim$ 2.0 \msun, they are fairly common and are of great interest,
since they allow us to understand the role of the stellar mass in the formation and characteristics of planetary systems. \newline \indent  
So far, we have discovered seven sub-stellar objects around six giant stars. Interestingly, all of these host stars are intermediate-mass stars, despite 
the fact that $\sim$ 25\% of our targets are low-mass stars. This result confirms that the frequency of planets increases with the stellar mass (e.g. 
Bowler et al. \cite{BOW10}; Johnson et al. \cite{JOHN10}; Reffert et al. \cite{REF15}). A detailed statistical analysis of the occurrence rate from the 
EXPRESS program will be presented soon (Jones et al. in preparation). \newline \indent
Finally, we present new radial velocity epochs of the giant star HIP\,67851, confirming the presence of an outer planet in the system. We obtained 
the following orbital parameter for HIP\,67851\,{\it c}: P = 2131.8 $\pm$ 43.5 d, m$_c$\,sin$i$ = 6.0 $\pm$ 0.7 \mjup and $e$ = 0.17 $\pm$ 0.04. 
Apart from HIP\,67851 there are eight giant stars (log$g$ $\lesssim$ 3.6) that are known to host multi-planet 
systems\footnote{The outer planet around HD\,47536 (Setiawan et al. \cite{SET08}) was shown likely not to be real (Soto et al. \cite{SOT15}). 
Also, the proposed system around BD+202457 (Niedzielski et al. \cite{NIE09}) was shown to be dynamically unstable (see Horner et al.  \cite{HOR14})}, 
namely 24 Sextantis and HD\,200964 (Johnson et al. \cite{JOHN11}), HD\,4732 (Sato et al. \cite{SAT13}), Kepler 56 (Huber et al. \cite{HUB13}), 
Kepler 432 (Quinn et al. \cite{QUI15}), Kepler 391 (Rowe et al. \cite{ROW14}), $\eta$ Ceti (Trifonov et al. \cite{TRI14}) and TYC 1422-614-1 
(Niedzielski et al. \cite{NIE15}). Interestingly, $\eta$ Ceti is the only one that is located in the clump region, although according to
Trifonov et al. (\cite{TRI14}) it is most likely a RGB star instead of a HB star.
The rest of them are very close to the base of the RGB phase, therefore they still have relatively small radii.
Moreover, Kepler 56\,{\it b}, Kepler 391\,{\it b}, Kepler 432\,{\it b}, HIP\,67851\,{\it b} and TYC-1422-614-1\,{\it b} are located in close-in orbits 
($a <$ 0.7 AU), a region where planets are rare around post-MS stars. This observational result suggests that systems with short period planets do exist 
around slightly evolved stars, but they are destroyed by the stellar envelope during the late stage of the RGB phase, when the stellar radius grows 
sufficiently large (R$_\star$ $\sim$ $a$). 
This could explain why these kind of systems are not found around HB giants and therefore why close-in planets are not found around such evolved stars. 
However, the stellar mass might be playing an important role on shaping the inner regions of planetary systems, particularly since
the aforementioned stars are on average less massive than the bulk of planet-hosting giant stars.

\begin{acknowledgements}
M.J. acknowledges financial support from Fondecyt project \#3140607 and FONDEF project CA13I10203.
J.J. and P.R. acknowledge funding by the CATA-Basal grant (PB06, Conicyt).
P.R. acknowledges support from Fondecyt project \#1120299. 
We acknowledge the anonymous referee for very useful comments, that helped to improve the quality of this work.
This research has made use of the SIMBAD database and the VizieR catalogue access tool, operated at CDS, Strasbourg, France. 
\end{acknowledgements}

\Online

\begin{appendix} 

\section{Radial velocity tables.}

\begin{table}
\centering
\caption{CHIRON radial velocity variations of HIP\,65891\label{chiron_HIP65891_vels}}
\begin{tabular}{lrr}
\hline\hline \vspace{-0.3cm} \\
JD\,-\,2450000        & RV & error  \\
     \,     &  (\ms)   &   (\ms) \\
\hline \vspace{-0.3cm} \\
  6523.4823  & -45.9 &  4.4 \\
  6536.4923  & -66.1 &  5.9 \\
  6654.7915  & -42.3 &  4.8 \\
  6670.8031  & -50.2 &  4.4 \\
  6673.8555  & -47.0 &  4.9 \\
  6678.8188  & -39.1 &  4.8 \\
  6689.8292  & -26.3 &  5.1 \\
  6708.8389  & -28.1 &  4.9 \\
  6722.6806  & -42.1 &  4.3 \\
  6736.6035  & -32.7 &  4.6 \\
  6752.6675  & -28.3 &  5.0 \\
  6769.6338  & -13.1 &  4.6 \\
  6785.6895  & -16.9 &  4.4 \\
  6809.5599  &  -8.1 &  4.4 \\
  6824.5449  &  -6.0 &  4.8 \\
  6857.4830  &  12.5 &  4.7 \\
  6879.5110  &  11.0 &  5.1 \\
  6880.4898  &   8.6 &  5.0 \\
  6888.4805  &  12.6 &  4.6 \\
  6898.4812  &  15.9 &  5.6 \\
  7018.8291  &  44.1 &  4.9 \\
  7028.8280  &  73.2 & 12.9 \\
  7038.8417  &  73.4 &  4.3 \\
  7044.8105  &  80.5 &  3.8 \\
  7046.7597  &  87.6 &  4.2 \\
  7050.7543  &  72.6 &  4.1 \\
\hline \vspace{-0.3cm} \\
\hline\hline
\end{tabular}
\end{table}

\begin{table}
\centering
\caption{FEROS radial velocity variations of HIP\,65891\label{feros_HIP65891_vels}}
\begin{tabular}{lrr}
\hline\hline  \vspace{-0.3cm} \\
JD\,-\,2450000        & RV & error  \\
          &  (\ms)   &   (\ms) \\
\hline \vspace{-0.3cm} \\
5317.6274  & -37.7 & 5.4 \\
5379.5439  & -33.5 & 5.3 \\
5428.5059  & -53.1 & 4.3 \\
5744.5713  &   1.4 & 6.6 \\
5786.5649  &  -1.8 & 5.1 \\
6047.5977  &  90.0 & 6.6 \\
6056.5850  &  71.7 & 5.0 \\
6066.5972  &  69.5 & 5.8 \\
6099.5825  &  57.9 & 5.9 \\
6110.5620  &  65.1 & 5.6 \\
6140.6284  &  37.6 & 4.8 \\
6160.5493  &  54.6 & 5.0 \\
6321.7749  &   5.5 & 7.2 \\
6331.8145  & -13.0 & 8.5 \\
6342.7432  &   0.3 & 7.4 \\
6412.6270  & -20.1 & 5.2 \\
6412.7261  & -35.9 & 4.1 \\
6412.7520  & -42.9 & 4.2 \\
6412.8042  & -39.0 & 4.0 \\
6431.5796  & -55.8 & 6.4 \\
6431.6436  & -55.7 & 6.1 \\
6431.6802  & -41.4 & 5.8 \\
6431.7441  & -43.4 & 6.9 \\
\hline\hline
\end{tabular}
\end{table}

\begin{table}
\centering
\caption{CHIRON radial velocity variations of HIP\,107773\label{chiron_HIP107773_vels}}
\begin{tabular}{lrr}
\hline\hline  \vspace{-0.3cm} \\
JD\,-\,2450000        & RV & error  \\
          &  (\ms)   &   (\ms) \\
\hline \vspace{-0.3cm} \\
  6223.5346  & -34.0  & 5.4 \\
  6230.5344  & -27.5  & 5.6 \\
  6241.5757  & -26.3  & 9.0 \\
  6248.5202  &  -8.1  & 7.9 \\
  6480.8846  & -33.5  & 4.6 \\
  6485.8048  & -48.7  & 5.5 \\
  6507.7828  & -43.8  & 22.7 \\
  6533.7565  &   6.9  & 5.1 \\
  6552.5484  &  24.4  & 7.8 \\
  6556.5333  &  54.7  & 7.2 \\
  6563.4964  &  51.5  & 7.6 \\
  6569.5599  &  38.7  & 7.1 \\
  6577.5286  &  27.4  & 6.6 \\
  6581.5111  &  47.6  & 6.4 \\
  6790.7998  & -45.5  & 5.0 \\
  6805.8595  & -35.1  & 4.8 \\
  6822.8399  &  32.9  & 6.7 \\
  6838.8336  &  31.1  & 5.4 \\
  6855.7274  &  21.0  & 5.1 \\
  6879.6707  &  36.1  & 4.7 \\
  6882.7526  &  23.1  & 5.0 \\
  6892.6339  &  19.3  & 5.9 \\
  6906.6214  &   5.8  & 5.9 \\
  6915.5634  & -28.3  & 5.0 \\
  6922.6643  & -35.6  & 4.8 \\
  6931.5519  & -62.2  & 5.2 \\
  6936.6461  & -30.1  & 5.5 \\
  6939.5186  & -31.1  & 5.3 \\
  6943.4909  & -23.7  & 5.4 \\
  6949.5001  &  -9.0  & 5.0 \\
  6951.4918  & -12.2  & 5.3 \\
  6957.5240  &   1.9  & 8.4 \\
  6966.5438  &   8.8  & 4.2 \\
  6975.5073  &   9.0  & 5.4 \\
  6983.5068  &  53.2  & 5.4 \\
  6991.5867  &  41.5  & 6.6 \\
\hline \hline
\end{tabular}
\end{table}

\begin{table}
\centering
\caption{FEROS radial velocity variations of HIP\,107773\label{feros_HIP107773_vels}}
\begin{tabular}{lrr}
\hline \hline  \vspace{-0.3cm} \\
JD\,-\,2450000        & RV & error  \\
          &  (\ms)   &   (\ms) \\
\hline \vspace{-0.3cm} \\
5317.8306  & -43.5 & 6.0 \\
5336.9233  & -54.8 & 5.1 \\
5366.9185  & -43.4 & 7.4 \\
5379.8804  & -48.1 & 4.4 \\
5428.7285  &  25.2 & 5.0 \\
5457.6816  & -31.1 & 4.8 \\
5729.8501  &   3.2 &10.2 \\
5729.8511  &   5.2 & 9.8 \\
5786.8003  & -45.2 & 3.2 \\
5786.8018  & -45.9 & 3.9 \\
5793.7930  & -17.1 & 5.3 \\
5793.7944  & -11.7 & 5.9 \\
6056.8149  & -35.6 & 5.1 \\
6110.8545  &  32.0 & 4.1 \\
6160.7500  &  21.5 & 4.5 \\
6230.5386  &   2.4 & 5.0 \\
6230.5396  &  -3.3 & 4.4 \\
6241.5571  &  26.7 & 6.6 \\
6241.5586  &  23.9 & 6.1 \\
6251.5811  &  51.1  &5.8 \\
6412.8145  &  51.4  &5.5 \\
6431.8267  &  65.5  &6.6 \\
6472.8037  &   9.6  &4.8 \\
6472.8442  &   5.3  &4.8 \\
6472.8892  & -17.1  &4.7 \\
6565.6064  &  76.4  &5.1 \\
6605.6260  &  43.2  &6.5 \\
\hline\hline
\end{tabular}
\end{table}

\begin{table}
\centering
\caption{CHIRON radial velocity variations of HIP\,67851\label{feros_HIP67851_vels}}
\begin{tabular}{lrr}
\hline \hline  \vspace{-0.3cm} \\
JD\,-\,2450000        & RV & error  \\
          &  (\ms)   &   (\ms) \\
\hline \vspace{-0.3cm} \\
  6299.8433 &   32.9  & 4.3 \\
  6307.8349 &   51.0  & 4.3 \\
  6312.8003 &   57.4  & 4.4 \\
  6317.7840 &   62.6  & 4.5 \\
  6331.8972 &  103.1  & 4.6 \\
  6338.8048 &  118.5  & 4.1 \\
  6374.7829 &   15.6  & 4.0 \\
  6384.6852 &   17.3  & 4.4 \\
  6385.6451 &   24.1  & 4.3 \\
  6411.5294 &   54.3  & 4.7 \\
  6422.6922 &   92.1  & 4.4 \\
  6426.5948 &   91.5  & 4.3 \\
  6476.6536 &   -7.9  & 3.7 \\
  6523.4691 &   62.5  & 4.0 \\
  6533.4833 &   48.6  & 5.1 \\
  6539.4800 &   40.3  & 9.4 \\
  6654.8190 &  -60.1  & 6.2 \\
  6656.8146 &  -47.7  & 5.7 \\
  6659.8055 &  -35.9  & 4.5 \\
  6666.8582 &  -12.8  & 5.0 \\
  6670.8156 &   13.2  & 4.9 \\
  6670.8601 &   -0.0  & 5.1 \\
  6676.7602 &   19.6  & 5.0 \\
  6683.8870 &   40.3  & 5.4 \\
  6690.8914 &   44.1  & 6.0 \\
  6708.8282 &   27.3  & 4.5 \\
  6722.6914 &  -22.3  & 4.3 \\
  6736.6157 &  -68.8  & 5.2 \\
  6752.6791 &  -45.7  & 4.5 \\
  6767.5400 &  -18.3  & 4.5 \\
  6783.7508 &    7.8  & 5.4 \\
  6798.6158 &    1.6  & 5.6 \\
  6819.5626 &  -69.7  & 6.5 \\
  6823.5774 &  -64.4  & 5.4 \\
  6840.5507 &  -61.8  & 4.3 \\
  7012.8699 & -108.0  & 6.3 \\
  7014.8328 & -100.3  & 5.5 \\
  7021.8543 &  -76.4  & 5.1 \\
  7029.8454 &  -51.5  & 4.6 \\
  7040.8343 &   -6.6  & 5.3 \\
  7044.7997 &  -12.8  & 5.3 \\
  7045.7595 &  -12.4  & 5.2 \\
  7060.8255 &   -7.5  & 5.3 \\
  7075.7198 &  -54.9  & 5.1 \\
  7089.6592 &  -80.0  & 4.7 \\
\hline\hline
\end{tabular}
\end{table}

\begin{table}
\centering
\caption{FEROS radial velocity variations of HIP\,67851\label{chiron_HIP67851_vels}}
\begin{tabular}{lrr}
\hline \hline  \vspace{-0.3cm} \\
JD\,-\,2450000        & RV & error  \\
          &  (\ms)   &   (\ms) \\
\hline \vspace{-0.3cm} \\
3072.8628  & -41.5 &20.0 \\
3072.8843  & -45.0 &20.0 \\
5317.6470  & -90.7 & 4.5 \\
5379.6509  & -15.4 & 5.0 \\
5729.6318  &  68.9 &11.1 \\
5744.6021  &  13.1 & 6.6 \\
6047.6216  &  24.4 & 7.0 \\
6056.6094  &  31.9 & 5.9 \\
6066.6245  &  54.0 & 6.3 \\
6099.6104  &   6.3 & 6.1 \\
6110.5845  & -14.0 & 5.3 \\
6140.6074  &  11.3 & 4.7 \\
6321.7993  &  22.7 & 8.4 \\
6331.7339  &  37.2 & 8.3 \\
6331.8057  &  31.3 & 7.3 \\
6342.7798  &  69.8 & 7.3 \\
6412.5737  &  20.7 & 5.4 \\
6431.5400  &  33.3 & 6.4 \\
6431.5869  &  39.8 & 6.8 \\
6431.6401  &  36.8 & 7.2 \\
6431.6855  &  35.6 & 6.3 \\
6431.7495  &  40.7 & 7.3 \\
6472.6060  & -59.7 & 5.2 \\
6472.6099  & -60.5 & 5.5 \\
6472.6274  & -57.9 & 5.9 \\
6472.6431  & -67.1 & 4.4 \\
6472.6572  & -63.1 & 4.3 \\
6472.7119  & -63.0 & 5.7 \\
\hline\hline
\end{tabular}
\end{table}

\end{appendix}


\begin{thebibliography}{}
\bibitem[1996]{BAR96} Baranne, A., Queloz, D., Mayor, M. et al. 1996, \aap, 119, 373
\bibitem[2003]{BEV03} Bevington, P. R. \& Robinson, D. K. 2003. Data Reduction and Error Analysis for the Physical Sciences, 3rd Ed. (McGraw-Hill) 
\bibitem[2010]{BOR10} Borucki, W. J., Koch, D., Basri, G. et al. 2010, Science, 327, 977
\bibitem[2010]{BOW10} Bowler, B. P., Johnson, J. A., Marcy, G. W. et al. 2010, \aap, 709, 396 
\bibitem[1996]{BUT96} Butler, R. P., Marcy, G. W., Williams, E. et al. 1996, PASP, 108, 500
\bibitem[2002]{FRI02} Frink, S., Mitchell, D. S., Quirrenbach, A. et al. 2002, \apj, 576, 478
\bibitem[2006]{HEK06} Hekker, S., Reffert, S., Quirrenbach, A. et al. 2006, A\&A, 454, 943
\bibitem[2010]{HOW10} Howard, A. W., Johnson, J. A., Marcy, G. W. et al. 2010, \apj, 721, 1467
\bibitem[2014]{HOR14} Horner, J., Wittenmyer, R. A., Hinse, T. C. \& Marshall, J. P. 2014, MNRAS, 439, 1176
\bibitem[2013]{HUB13} Huber, D., Carter, J. A., Barbieri, M. et al. 2013, Science, 342, 331
\bibitem[2008]{JEN08} Jenkins, J. S., Jones, H. R. A., Pavlenko, Y. et al. 2008, \aap, 485, 571
\bibitem[2011]{JEN11} Jenkins, J. S., Murgas, F., Rojo, P. et al. 2011, A\&A, 531, 8
\bibitem[2006]{JOH06} Johnson, J. A., Marcy, G. W., Fischer, D. A. et al. 2006, \apj, 652, 1724
\bibitem[2010]{JOHN10} Johnson, J. A., Aller, K. M., Howard, A. W. \&  Krupp, J. R. 2010, PASP, 122, 905
\bibitem[2011]{JOHN11} Johnson, J. A., 	Johnson, John A., Payne, M., Howard, A. W. et al. AJ, 141, 16
\bibitem[2011]{JON11} Jones, M. I., Jenkins, J. S., Rojo, P. \& Melo, C. H. F. 2011, A\&A, 536, 71
\bibitem[2013]{JON13} Jones, M. I., Jenkins, J. S., Rojo, P., Melo, C. H. F. \& Bluhm, P. 2013, A\&A, 556, 78
\bibitem[2014]{JONJEN14} Jones, M. I \& Jenkins, J. S. 2014, A\&A, 562, 129
\bibitem[2014]{JON14} Jones, M. I., Jenkins, J. S., Bluhm, P., Rojo, P. \& Melo, C. H. F. 2014, A\&A, 566, 113
\bibitem[2015]{JON15} Jones, M. I., Jenkins, J. S., Rojo, P., Melo, C. H. F. \& Bluhm, P. 2015, A\&A, 573, 3
\bibitem[1999]{KAU99} Kaufer, A., Stahl, O., Tubbesing, S. et al. 1999, The Messenger 95, 8
\bibitem[1995]{KJE95} Kjeldsen, H. \& Bedding, T. R. 1995, A\&A, 293, 87 
\bibitem[2011]{LLO11} Lloyd, J. P. 2011, \apj, 739, 49
\bibitem[2013]{LLO13} Lloyd, J. P. 2013, \apj, 774, 2    
\bibitem[1995]{MAY95} Mayor, M. \& Queloz, D. 1995, Nature, 378, 355
\bibitem[2009]{MES09} Meschiari, S., Wolf, A. S., Rivera, E. et al. 2009, \pasj, 121, 1016
\bibitem[2009]{NIE09} Niedzielski, A., Nowak, G., Adamów, M. \& Wolszczan, A. 2009, \apj, 707, 768
\bibitem[2015]{NIE15} Niedzielski, A., Villaver, E., Adamów, M. et al. 2015, A\&A, 573, 36 
\bibitem[2015]{QUI15} Quinn, S. N., White, T. R., Latham, D. W. et al. 2015, ApJ, 803, 49          
\bibitem[2015]{REF15} Reffert, S., Bergmann, C., Quirrenbach, A. et al. 2015, A\&A, 574, 116  
\bibitem[2014]{ROW14} Rowe, J. F., Bryson, S. T., Marcy, G. W. et al. 2014, \apj, 784, 45
\bibitem[2000]{SAL00} Salasnich, B., Girardi, L., Weiss, A. \& Chiosi, C. 2000, A\&A, 361, 1023
\bibitem[2005]{SAT05} Sato, B., Eiji, K., Yoichi, T. et al. 2005, PASJ, 57, 97
\bibitem[2013]{SAT13} Sato, B., Omiya, M., Wittenmyer, R. A., et al. 2013, \apj, 762, 9
\bibitem[2008]{SET08} Setiawan, J., Weise, P., Henning, Th. et al. 2008, arXiv:0704.2145 
\bibitem[1982]{SCA82} Scargle, J. D. 1982, \apj, 263, 835
\bibitem[2013]{SCH13} Schlaufman, K. \& Winn, J. 2013, \apj, 772, 143
\bibitem[2015]{SOT15} Soto, M., Jenkins, J. S. \& Jones, M. I. 2015, arXiv:1505.04796
\bibitem[2013]{TOK13} Tokovinin, A., Fischer, D. A., Bonati, M. et al. 2013, PASP, 125, 1336
\bibitem[2014]{TRI14} Trifonov, T., Reffert, S., Tan, X., Lee M. H., \& Quirrenbach, A. 2014, A\&A, 568, 64
\bibitem[2007]{VAN07} Van Leeuwen, F. 2007, A\&A, 474, 653
\bibitem[2009]{WINN09} Winn, J. N.,  Johnson, J. A., Albrecht, S. et al. 2009, ApJ, 703, 99
\bibitem[2011]{WIT11} Wittenmyer, R. A., Endl, M., Wang, L. et al. 2011, ApJ, 743, 184
\bibitem[2015]{WIT15} Wittenmyer, R. A., Wang, L., Liu, F. et al. 2015, \apj, 800, 74 
\bibitem[1992]{WOL92} Wolszczan, A. \& Frail, D. A. 1992, Nature, 355, 14 1992, Nature, 355, 1455

\end{thebibliography}
\end{document}